\documentclass[manuscript=false]{acmart} 
\usepackage{graphicx} 
\usepackage{adjustbox}
\usepackage[nolist,smaller]{acronym}
\usepackage{xcolor} 
\usepackage{tikz} 
\usepackage{wrapfig}
\usepackage{listings}
\usepackage{multirow}
\usepackage{dirtytalk}
\usepackage{amsmath}
\usepackage{tabularx}
\usepackage{array}
\usepackage{booktabs}
\newcommand{\etal}{\textit{et al.}}

\definecolor{figureborder}{HTML}{C8C8C8}
\definecolor{ogblue}{HTML}{518CE5}
\definecolor{oggreen}{HTML}{7DBA4C}
\definecolor{ogred}{HTML}{A22025}

\definecolor{condred}{HTML}{B1001C}
\definecolor{condorange}{HTML}{B34400}
\definecolor{taskblue}{HTML}{003776}
\definecolor{taskpurple}{HTML}{310067}

\title{\platforms{}: AI-Enhanced Interactive Narratives for Programming Education}

\date{June 2023}

\newcommand{\platform}[1]{EDBook}
\newcommand{\platforms}[1]{EDBooks}
\newcommand{\instructor}[1]{Dr. Ed}

\newcommand{\branch}{\includegraphics[height=2ex]{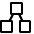}}

\newcommand{\anonymized}[1]{\textit{\textcolor{darkgray}{#1}}}

\newcommand{\condedbook}[1]{$\mathcal{C}$\textsubscript{\textcolor{condred}{\tiny{EDBook}}}}
\newcommand{\condbaseline}[1]{$\mathcal{C}$\textsubscript{\textcolor{condorange}{\tiny{Web+GPT}}}}
\newcommand{\taskcache}[1]{$\mathcal{T}$\textsubscript{\textcolor{taskpurple}{\tiny{cache}}}}
\newcommand{\taskscheme}[1]{$\mathcal{T}$\textsubscript{\textcolor{taskblue}{\tiny{scheme}}}}
\newcommand{\participantquote}[2]{\begin{quote}``\textit{#1}'' (#2)\end{quote}}
\newcommand{\inlineparticipantquote}[2]{``\textit{#1}'' (#2)}

\begin{acronym}
    \acro{CUI}{Conversational User Interface}
    \acro{IPA}{Intelligent Personal Assistant}
    \acro{IDE}{Integrated Development Environment}
    \acro{CQA}{Community Question-Answering}
    \acro{API}{Application Programming Interface}
    \acro{RSSE}{Recommendation System in Software Engineering}
    \acro{AJAX}{Asynchronous JavaScript and XML}
    \acro{JSON}{JavaScript Object Notation}
    \acro{HTML}{HyperText Markup Language}
    \acro{URL}{Uniform Resource Locator}
    \acro{VCS}{Version Control System}
    \acro{AI}{Artificial Intelligence}
    \acro{SHOCK}{SHOCK}
    \acro{ROC}{ROC}
    \acro{HCI}{Human-Computer Interaction}
    \acro{SE}{Software Engineering}
    \acro{VR}{Virtual Reality}
    \acro{AR}{Augmented Reality}
    \acro{MR}{Mixed Reality}
    \acro{HMD}{Head-Mounted Display}
    \acro{IE}{Internet Explorer}
    \acro{IEEE}{Institute of Electrical and Electronics}
    \acro{CHI}{the \acs{ACM} Conference on Human Factors in Computing Systems}
    \acro{UIST}{the \acs{ACM} Symposium on User Interface Software and Technology}
    \acro{CSCW}{the \acs{ACM} Conference on Computer-Supported Cooperative Work and Social Computing}
    \acro{VL/HCC}{the \acs{IEEE} Symposium on Visual Languages and Human-Centric Computing}
    \acro{NSF}{National Science Foundation}
    \acro{PI}{PI}
    \acro{co-PI}{co-PI}
    \acro{MOOC}{Massive Open Online Course}
    \acro{IVA}{Intelligent Virtual Assistant}
    \acro{OT}{Operational Transformation}
    \acro{L@S}{The ACM Conference on Learning at Scale}
    \acro{WYSIWIS}{What You See Is What I See}
    \acro{IIS}{Information and Intelligent Systems}
    \acro{VPL}{Visual Programming Language}
    \acro{2D}{two-dimensional}
    \acro{3D}{three-dimensional}
    \acro{CAD}{Computer-Aided Design}
    \acro{FRP}{Functional Reactive Programming}
    \acro{FSM}{Finite State Machine}
    \acro{VRML}{Virtual Reality Modeling Language}
    \acro{UI}{User Interface}
    \acro{GUI}{Graphical User Interface}
    \acro{WIMP}{Window Icon Mouse Pointer}
    \acro{EUP}{End User Programming}
    \acro{CAREER}{CAREER}
    \acro{HTC}{High Tech Computer Corporation}
    \acro{IoT}{Internet of Things}
    \acro{CS}{Computer Science}
    \acro{ACM}{the Association for Computing Machinery}
    \acro{SIGCSE}{the \acs{ACM} Special Interest Group on Computer Science Education}
    \acro{AIED}{Artificial Intelligence in Education}
    \acro{LAK}{Learning Analytics and Knowledge}
    \acro{SI}{School of Information}
    \acro{UMSI}{the University of Michigan School of Information}
    \acro{MOOC}{Massive Open Online Course}
    \acro{AST}{Abstract Syntax Tree}
    \acro{LLM}{Large Language Model}
    \acro{tf-idf}{term frequency–inverse document frequency}
    \acro{T-SNE}{T-Distributed Stochastic Neighbor Embedding}
    \acro{PCA}{Principal Component Analysis}
    \acro{CFG}{Control Flow Graph}
    \acro{ACFG}{Aggregated Congrol Flow Graph}
    \acro{CRDT}{Conflict-free Replicated Data Type}
    \acro{SDL}{Self-Directed Learning}
    \acro{IRB}{Institutional Review Board}
\end{acronym}

\begin{document}

\author{Steve Oney}
\orcid{0000-0002-5823-1499}
\affiliation{
    \institution{University of Michigan}
    \city{Ann Arbor}
    \state{Michigan}
    \country{USA}
}
\email{soney@umich.edu}
 
\author{Yue Shen}
\affiliation{
    \institution{University of Michigan}
    \city{Ann Arbor}
    \state{Michigan}
    \country{USA}
}
\email{yuesh@umich.edu}

\author{Fei Wu}
\affiliation{
    \institution{University of Michigan}
    \city{Ann Arbor}
    \state{Michigan}
    \country{USA}
}
\email{feiyuwu@umich.edu}

\author{Young Suh Hong}
\affiliation{
    \institution{University of Michigan}
    \city{Ann Arbor}
    \state{Michigan}
    \country{USA}
}
\email{hngchris@umich.edu}

\author{Ziang Wang}
\affiliation{
    \institution{University of Michigan}
    \city{Ann Arbor}
    \state{Michigan}
    \country{USA}
}
\email{ziangw@umich.edu}

\author{Yamini Khajekar}
\affiliation{
    \institution{Indian Institute of Technology}
    \city{Delhi}
    \state{Delhi}
    \country{India}
}
\email{yaminikhajekar.iitd@gmail.com}

\author{Jiacheng Zhang}
\affiliation{
    \institution{University of Michigan}
    \city{Ann Arbor}
    \state{Michigan}
    \country{USA}
}
\email{jiache@umich.edu}

\author{April Yi Wang}
\affiliation{
    \institution{ETH Zürich}
    \city{Zürich}
    \country{Switzerland}
}
\email{april.wang@inf.ethz.ch}

\begin{abstract}

\acfp{LLM} have shown the potential to be valuable teaching tools, with the potential of giving every student a personalized tutor.
However, one challenge with using \acp{LLM} to learn new concepts is that when learning a topic in an unfamiliar domain, it can be difficult to know what questions to ask.
Further, language models do not always encourage ``active learning'' where students can test and assess their understanding.
In this paper, we propose ways to combine large language models with ``traditional'' learning materials (like e-books) to give readers the benefits of working with \acp{LLM} (the ability to ask personally interesting questions and receive personalized answers) with the benefits of a traditional e-book (having a structure and content that is pedagogically sound).
This work shows one way that \acp{LLM} have the potential to improve learning materials and make personalized programming education more accessible to a broader audience.\end{abstract}
\maketitle

\section{Introduction}\emph{Dialogic learning} emphasizes interactive, learner-centered approaches where students actively co-construct knowledge through dialogue \cite{wegerif2007dialogic}.
Dialogic learning can be used for a wide variety of subjects, including computing education.
Prior work has shown many benefits of dialogic learning \cite{teo2019teaching}, including increasing students' engagement, encouraging active learning \cite{wood2018teacher}, critical thinking \cite{frijters2008effects}, and providing personalized learning experiences.
However, providing dialogic learning at scale remains challenging \cite{resnick2018next}.
Dialogic learning requires that instructors actively engage students, respond to their questions, and provide meaningful feedback.
This is especially important in programming education, which requires active hands-on practice and feedback to build skills.
Traditional classrooms can offer effective teaching and support but often lack enough instructors to for personalized feedback.
Non-traditional environments like online courses and \ac{SDL} can provide scalable instruction but may lack personalized guidance and meaningful practice.

Although dialogic learning is primarily focused on synchronous, interactive educational experiences, the principles can also be applied to static educational content.
For example, Daniel Friedman authored a series of books that teach computing concepts through a dialog between a learner and an instructor~\cite{friedman1995little,friedman1995seasoned}.
However, prior approaches \cite{friedman1995little, wang2019designing} using the question-and-answer format to write programming learning materials have limitations in providing truly adaptive and interactive content tailored to diverse learners' needs.
Thus, the question that motivates this research is: \textbf{how can we enable high-quality, personalized, and interactive dialogic learning experiences at scale?}
We are particularly focused on supporting dialogic learning for \emph{asynchronous} programming education, which can include \ac{SDL} or assignments that are part of a traditional course.

Recently, the advent of \acp{LLM} such as ChatGPT has opened new possibilities for developing engaging and personalized educational content that can provide responsive guidance at scale~\cite{heaven2023chatgpt,lin2023exploring}.
\acp{LLM} can accurately respond to a wide variety of natural language questions and requests.
However, one challenge with using open-ended \acp{LLM} to learn new concepts is that it can be difficult for learners to know what questions to ask to effectively build understanding in an unfamiliar domain.
Although many \emph{task-oriented} \ac{LLM} extensions (such as chatbots for booking flights) can naturally elicit information from users (e.g., departure airport, destination, etc.), they do not have a way to ensure readers understand educational content.
Ideally, learners should have the flexibility to explore and ask questions on their own while achieving larger learning goals, which could be specified by instructors and field experts.
This highlights a key challenge: \textbf{how can we enable programming learners to benefit from interactive \acp{LLM} while aligning interactions with pedagogical goals?}


\begin{figure}[t]
    \centering
    \includegraphics[width=\textwidth]{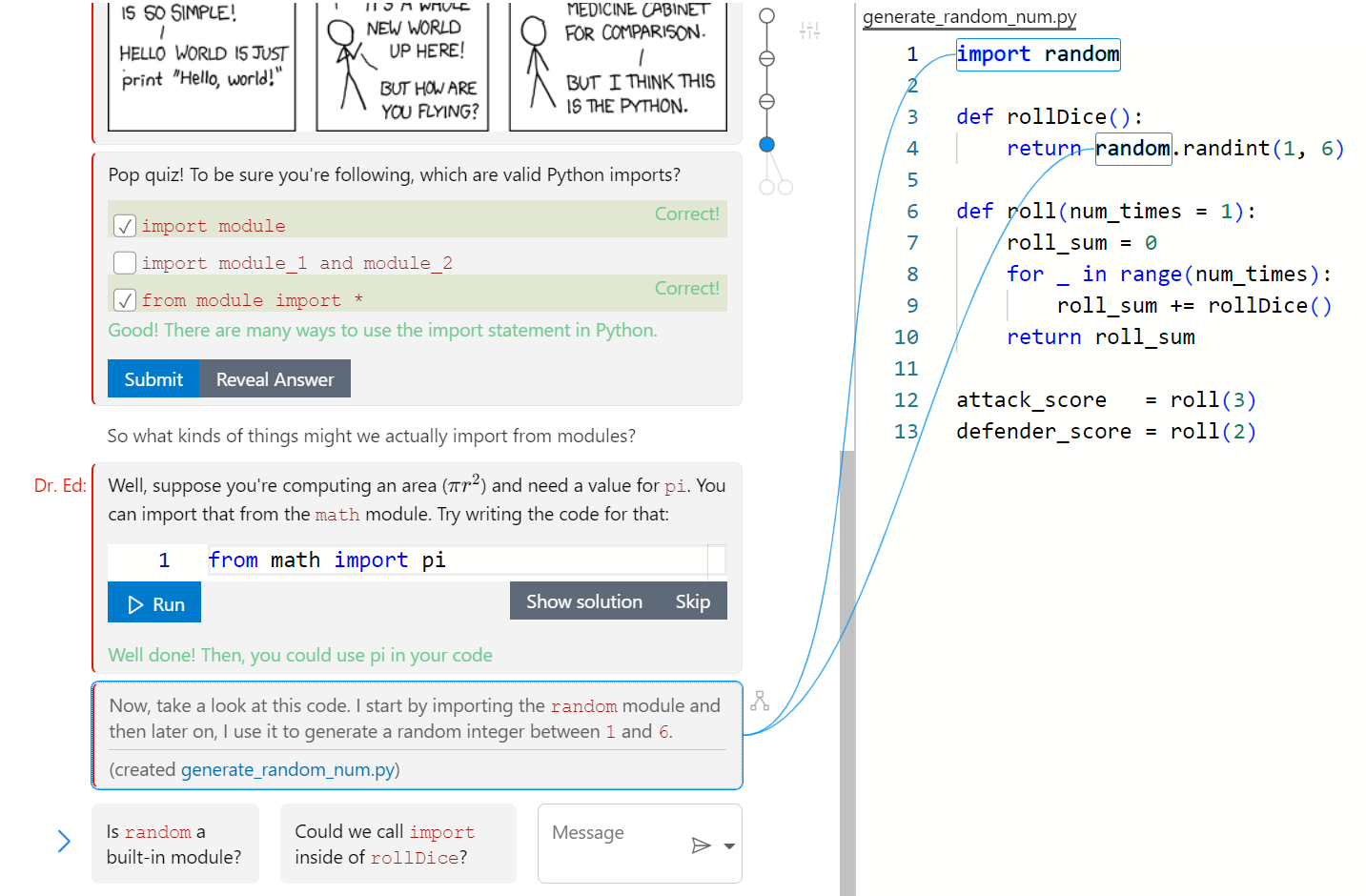}
    \caption{An illustration of \platform{} content that describes the basics of Python modules.
    Readers learn through an interactive dialog (left column) that includes a natural language didactic dialog, multiple choice questions, and code writing questions.
    With \platform{}, readers work through this narrative over time.
    At any point in the narrative, they can also interact with a context-aware \ac{LLM} for additional questions.
    These dialogs can also include larger code samples that are incrementally built and explained over the course of the narrative (right column).
    Deictic pointers (represented as blue lines) can link the narratives with specific parts of code.
    At the bottom of the leftmost column, the reader is given three options for how to proceed with the narrative: two pre-built questions (``is \texttt{random} a built-in module?'' and ``Could we call \texttt{import} inside of \texttt{rollDice}?'') and the option to query an \ac{LLM}.
    }
    \label{fig:edbook-main}
\end{figure}

In this paper, we propose a platform that addresses these challenges by enabling a limited form of dialogic learning.
We designed \textbf{\platform{}}\footnote{``\platform{}'' is short for \textbf{\underline{E}}lectronic \textbf{\underline{D}}ialogic \textbf{\underline{Book}}} a new platform for shareable, interactive educational content\footnote{Readers are invited to try an anonymized deployment of \platform{} at \url{https://edbook.net/}.}.
With \platform{}, content authors can write dialog trees---dialogs where messages can have multiple responses---that are aimed towards a learning goal.
These dialogs can involve any number of `participants' but typically contain at least an instructor persona who guides readers through didactic dialog and a persona representing the reader.
They can contain interactive code-writing questions to give readers hands-on practice, multiple-choice questions to help readers assess their understanding, larger code samples that are incrementally built and explained over the course of the dialog, and multimedia, as Figure~\ref{fig:edbook-main} shows.
Readers can interact with a pre-made dialog tree or can self-explore through open-ended interactions with a context-aware \ac{LLM}.
Readers are guided towards pre-determined learning goals, which are represented by a `target' point in the dialog tree---typically a leaf node that represents the end of the dialog.

\textbf{Contributions:}
The key innovation of this work is in enabling dialogic learning experiences that are open-ended yet goal-aligned.
By doing this, we can guide students towards learning goals while providing responsive guidance and reinforcing concepts through interactive practice.
To our knowledge, this represents the first system to combine these techniques for programming instruction.

The key contributions of this work are:
\begin{itemize}
    \item The design of \platform{}, the first platform to integrate pre-written dialogic narratives with open-ended context-informed \acp{LLM}.
    \platform{} also contains additional features (like code writing questions and incremental code building) specifically designed for programming education.
    \item Techniques to align open-ended \ac{LLM} interactions with larger learning goals, enabling learner agency while meeting pedagogical objectives. 
    \item A representation for branched dialogs, code examples, and learner state that maintains readability and understandability for learners.
    \item A study evaluating learning outcomes and qualitative experiences using \platform{} for programming education.
\end{itemize}

The \platform{} platform also includes several additional features that are not entirely novel but that are complimentary to its design, including
(1) interactive quiz questions (including code-writing questions and multiple-choice questions) that help students assess their understanding of the material,
(2) the ability to incrementally build up and explain large code samples over the course of a narrative, and
(3) deictic code pointers that tie narratives with the larger code samples they discuss.
Although these features are not novel independently, we show how they can be effectively integrated into the larger \platform{} platform and dialogic learning more generally.
Overall, our results highlight the promise of dialogic learning with \acp{LLM} for increased engagement, agency, and flexible learning tools that scale.
\section{Related Work}
Our work on the \platform{} platform builds on a rich body of prior work on \acp{LLM} in education, textbook platforms, techniques for connecting code with dialog, and educational narratives.

\subsection{Using \acf{AI} and \acfp{LLM} as Learning Tools}

Researchers and practitioners have recognized the value of \acf{AI} as a learning tool~\cite{heaven2023chatgpt,hirsh2023chatgpt}, particularly for \acf{SDL}~\cite{lin2023exploring}.
Prior work has integrated \ac{AI} and dialog systems into education in a variety of ways.
Conversational agents and chatbots have been used to support personalized learning experiences through natural dialog~\cite{han2021designing, li2022chatbot, luo2020code}.
Intelligent tutoring systems can adaptively provide hints and feedback to learners, similar to human tutors~\cite{jell2023towards}.
They can also be particularly useful for learning new natural languages~\cite{bibauw2022dialogue}.
These systems demonstrate the potential for \ac{AI} to enable responsive and adaptive education.

More recent work on \ac{AI} in education has also examined \acfp{LLM} in particular.
\acp{LLM} can answer student questions~\cite{hirsh2023chatgpt, joyner2023chatgpt,liffiton2023codehelp}, generate explanations~\cite{leinonen2023explanations}, and even solve many coding problems~\cite{kazemitabaar2023studying, reeves2023evaluating}. 
However, if not constrained properly, \acp{LLM} may produce misleading, incorrect, or educationally unproductive responses~\cite{guo2023six}.
Effectively aligning \acp{LLM} interactions with learning goals remains an open challenge.
%

Within this context, platforms like Graphologue~\cite{jiang2023graphologue} and Sensecape~\cite{suh2023sensecape} both explore ways to leverage \acp{LLM} to arrange learning materials.
However, Graphologue's design is oriented towards building a deep understanding of one particular concept by exploring multiple levels of abstraction for \ac{LLM} responses and Sensecape's design is oriented towards sensemaking.
Neither is focused on longer educational narratives, as the \platform{} platform is.
The features of Graphologue could be combined with the \platform{} platform to produce narratives that have more `local' detail throughout.

One feature of the \platform{} platform is that it can represent and help authors automatically generate dialogic content between any number of agents, based on descriptions of those agents.
Commercial tools like character.ai~\cite{CharacterAI} also allow users to create conversations with simulated agents.
Similarly, Markel \etal{} explored how \acp{LLM} can simulate students to help train instructors~\cite{markel2023gpteach}.
Unlike these platforms, \platform{} dialogs are designed to guide readers towards targeted learning goals.

\subsection{The Impact of \acp{LLM} on Programming Education}
Since modern \acp{LLM} like GPT-3 have proven to be capable of solving a variety of problems~\cite{bubeck2023sparks}---particularly for programming problems---researchers have recognized that \acp{LLM} are likely to have a profound effect on almost every aspect of programming education~\cite{becker2023programming}.
This includes how we assess understanding, how we generate practice exercises~\cite{lu2023readingquizmaker}, how we create programming tutorials~\cite{guo2023six}, and what is pedagogically important to teach (e.g., perhaps de-emphasizing low-level implementation work that can be automated).
Prior work has studied many aspects of the impact of \acp{LLM} on programming education.
Kazemitabaar \etal{} found that code generation tools can be a valuable learning tool in introductory programming education and that using code generation tools in the context of learning could reduce frustration without negatively impacting students' understanding~\cite{kazemitabaar2023studying}.
Sarsa \etal{} found that \acp{LLM} can also generate high-quality coding exercises that can be directly used by instructors in their courses~\cite{sarsa2022exercise}. 
Additionally, the code explanations produced by \acp{LLM} also demonstrate a high level of correctness, making it a promising tool with the potential to expedite tutoring processes for teaching assistants~\cite{leinonen2023explanations,sarsa2022exercise}.

For novice programming students, the incorporation of \acp{LLM} could potentially reframe their educational objectives. 
There arises a heightened emphasis on acquiring the proficiency to explain and communicate algorithms, as it directly influences the accuracy of code produced by \acp{LLM}~\cite{yujia2022science}. 
In addition, the significance of code debugging and extending should also be emphasized, given that \acp{LLM} is capable of generating basic code for learners~\cite{becker2023programming}. 
By leveraging the capabilities of \acp{LLM}, students can more easily understand error messages, identify mistakes, and accelerate their overall learning progress~\cite{leinonen2023error}.

The incorporation of \acp{LLM} into programming education also raises certain concerns. 
Code generated by \acp{LLM} can exhibit biased structures and comments \cite{pearce2022unsecure} and may also contain insecure vulnerabilities \cite{chen2021evaluating}. 
Furthermore, Becker \etal{} suggest that the coding style and approaches adopted by \acp{LLM} might prove too advanced for novice programmers to readily adopt \cite{becker2023programming}. 
In this case, \platform{} aims to create a semi-structured learning environment, enabling learners to engage with \acp{LLM} under appropriate, dialogue-style guidance. 
This approach is designed to mitigate the potential negative impacts that could be transferred to learners.


\subsection{Interactive Textbook Platforms}
There are several interactive textbook platforms, particularly for teaching programming. Runestone~\cite{ericson2020runestone}, a platform on which several widely-used programming textbooks are built, enables interactive content, including code-writing questions, multiple choice, and Parsons problems~\cite{denny2008evaluating} (where students drag and drop code blocks to arrange them in a correct order rather than writing code manually).
Jupyter Book~\cite{executable_books_community_2020_4539666} is another computational platform that empowers the creation of interactive textbooks.
It integrates narrative text, code execution, visualizations, and interactive widgets within the environment.
Al-Gahmi \etal{} emphasize the effectiveness of using Jupyter as a valuable tool in computing education within the classroom, leading to significant improvements in students' performances on assignments \cite{al-gahmi2022jupyter}.
OpenDSA \cite{shaffer2011opendsa} is an open-source platform designed to streamline the creation of interactive textbooks for computer science-related subjects.
It enables instructors to create visualizations, quizzes, and animations within the web-based book, enhancing the learning experience for students.
Prior studies indicate that many students prefer interactive textbooks over conventional text materials~\cite{smith2021modeling,tommy2016opendsa}. 
However, existing interactive textbook solutions have certain limitations.
Computational notebooks such as Jupyter Notebook might not be intuitive for novice learners~\cite{johson2020jupyterinclassroom}.
On the other hand, other web-based textbooks can lack support for complex code execution and note-taking~\cite{miller2012beyondpdf}.
Consequently, students need to use multiple platforms simultaneously.
%
The \platform{} platform is designed as an extension of the Visual Studio Code platform.
By integrating learning and coding within a single interface, \platform{} aims to help students engage with content and practice coding cohesively.

\subsection{Goal-Oriented Dialog Systems and State-Based Representations of Dialogs}
Goal-oriented dialog systems are dialog systems that give users agency and choice while still directing them towards a set goal.
Goal-oriented dialogue systems are commonly applied to chatbots~\cite{grudin2019chatbots,xie2022converse}, which are developed to provide real-time assistance to users in accomplishing specific tasks. 
Prior studies indicate that numerous goal-oriented dialogue chatbots struggle to meet users' expectations \cite{jain2018evaluating,zamora2017m}. 
This challenge arises due to the intricacies of automating certain tasks \cite{cranshaw2017calendar}, the complexities of dialogues \cite{grudin2019chatbots}, and the inherent difficulty in accurately identifying natural language content \cite{li2020conversation}.
Most systems are targeted towards goals that are specific and concrete, such as booking a flight or scheduling a meeting~\cite{muise2019planning,santos2022review}.
By contrast, notebooks in \platform{} have the goal of building conceptual understanding for readers.
To mitigate the challenges presented by more open-ended goal-oriented dialog systems, \platform{} adopts a tree-based scripted dialogue style for its notebooks to minimize dialogue variability. 
Additionally, the preset message options can also reduce potential interruptions during dialogues \cite{li2020conversation}.
In \platform{}, the learning objectives can be considered ``abstract tasks''. 
To aid readers in identifying their ``task progress'', \platform{} incorporates quizzes within its notebooks.
%
Many tools, including Dialogflow~\cite{Dialogflow} and Tae \etal{}'s work~\cite{kim2023cells}, use state-based representations of dialogs.
This feature supports users to go back to the previous chat boxes and modify any cell to their favor. 
Extensive research has also been done in predicting future utterances from textual data, such as example-based dialog systems~\cite{lee2006situation}.
Example-based dialog systems can generate system responses for user input based on stored data examples.


After we have considered possible visualized dialog representations from traditional spoken dialogs~\cite{dahl2005visualization}, we finally decide to use representations of dialogs that are both state-based and tree-based. Notably, our work is designed and implemented as a tool to help users in navigating programming instructional content. It serves the purpose of guiding users through the proposed learning path while also alerting them if their progression diverges from the initially intended learning trajectory. This is achieved by a seamless amalgamation of the state-based and tree-based representations. Moreover, the system enables users to embark on exploratory journeys, empowering them to attempt different paths within the instructional dialogue. If users acquire new insights during their exploration and decide to revise their approach, the system facilitates a fluid return to previous conversations.

Topic shifting~\cite{tang2019target}---shifting open-ended conversations towards a target topic---is also a promising approach.
However, existing topic-shifting approaches are designed for conceptually simple topics (single entities, like `football')~\cite{tang2019target} rather than complex learning goals.
Further, topic shifting relies on \ac{LLM}-generated content, which can be inaccurate or misaligned with learning goals.

\subsection{Connecting Code and Dialog}
chat.codes~\cite{oney2018creating} is a tool for discussing code.
Although the use case of \platform{} is different, there is overlap in the features of \platform{} and chat.codes.
Like chat.codes~\cite{oney2018creating}, \platform{} enables programmers to create deictic references between messages and specific regions of code.
However, \platform{}'s design of these deictic references differs from that of chat.codes in several important ways.
First, deictic references in chat.codes needed to be specified on the \emph{character} level through markdown annotations.
For example: ``\texttt{[these lines](code.js:L23-L42)}'' creates a deictic reference in chat.codes.
If the user hovered over ``these lines'', lines 23--42 in \texttt{code.js} would be highlighted.
In our experimentation with writing content, we found that we rarely needed to write fine-grained references.
Thus, deictic references in \platform{} are made at a \emph{cell} level---cells contain a list of references, rather than specific parts of text.
We found that this change simplified deictic pointers for both authors and readers.
For authors, \platform{} reduces the amount of syntax they need to learn and allows them to add deictic references through simpler point-and-click interactions\footnote{chat.codes includes \ac{UI} functionality for adding deictic references without typing out the syntax manually. Users can highlight a section of an un-sent message and then highlight a region of code to add a reference to the highlighted part of the message. However, we found that this interaction was not discoverable and could be difficult for authors, particularly when they made subsequent modifications.}.
For readers, \platform{}'s representation of deictic pointers allows them to see them automatically.
Unlike in chat.codes, where readers needed to actively hover over the relevant text to see pointers, \platform{} can always display deictic references for the selected cell(s).
Further, \platform{} visualizes a curved line between the selected cell(s) and the code they discuss, which helps form a clearer connection between the narrative and code.

Callisto~\cite{wang2020callisto} also contains deictic references, in the context of the Jupyter Notebook platform.
Callisto's deictic references are also coarser than those of chat.codes, with 
Like \platform{}, Callisto's references are made on a cell-level, where relevant code will be highlighted when one message is selected.
The primary distinction between Callisto and \platform{} lies in their approach to highlighting code for deictic references.
In Callisto \cite{wang2020callisto}, a highlight color is used to emphasize relevant content, including not only code but also notebook cells, markers, snapshots, and different versions. This highlight color strategy maintains a uniform experience across various content types, effectively highlighting the contextual discussion of selected messages.
In contrast, the \platform{} utilizes a curved line to connect relevant code and cells. This method is designed primarily for code referencing, which creates a more intuitive connection between code and narrative cells.

Colaroid~\cite{wang2023colaroid}, another extension for Visual Studio Code, also combines narrative text with iterative code changes for authoring coding tutorials.
Like \platform{}, Colaroid enables authors to explain larger pieces of code through a series of steps.
However, there are several key differences between Colaroid and \platform{}.
First, Colaroid does not contain any \emph{dialogic} features---readers cannot interrupt the narrative to query an \ac{LLM} and authors cannot build assessments (e.g., multiple-choice or coding questions) into their narratives.
Second, Colaroid does not include deictic references to better tie the narrative text with content.
Finally, \platform{} gives users more agency in choosing how to follow a given tutorial. Readers can select which response they want to give and can write their own parts of narratives.
However, Colaroid also has several features that \platform{} would benefit from---most notably, the ability to integrate with git and version control tools and the ability to re-play user interactions on a \ac{UI}.

Like the \platform{} platform, Torii~\cite{head2020composing} also allows code examples to be built up incrementally.
However, beyond this similarity, there are significant differences between the design and use cases for these two tools.
\platforms{} focus more on explanation, exploration, and learning through interactive dialog whereas Torii is designed to give readers clear walkthroughs of a single codebase.


\section{\platform{} Design}We divide our discussion of the \platform{} design into two.
First, we describe the core mechanics of \platform{}.
Then, we describe how complementary features, like interactive questions integrate with the \platform{} platform.
Finally, we conclude by describing the aspects of \platform{} that are novel and unique relative to prior work.

\subsection{Core Design and Novelty}

The core design of the \platform{} platform is in how it combines dialog trees with open-ended \ac{LLM} interactions.
The fundamental design questions are about balancing the ability for learners to explore while meeting concrete learning goals and how to represent dialogs in a way that is coherent and understandable.

\subsubsection{Balancing Free Exploration with Concrete Learning Goals}
\begin{figure}
    \centering
    \includegraphics[width=0.8\textwidth]{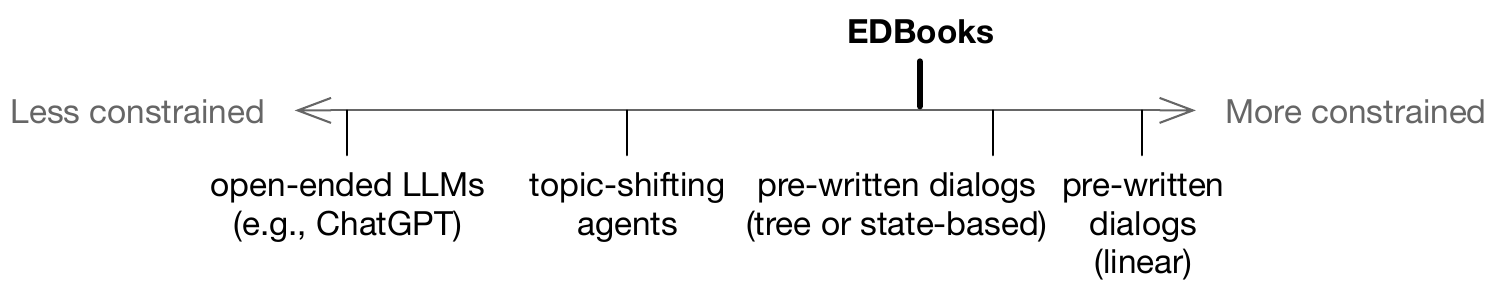}
    \caption{An illustration of different dialog system types along a continuum from less constrained input (left) to more (right).}
    \label{fig:llm_open_ended_design_space}
\end{figure}

One of the core design questions is how to balance concrete learning goals with open-ended exploration.
Figure~\ref{fig:llm_open_ended_design_space} illustrates where various dialog system techniques stand on a continuum of degree of constraint.
Less constrained systems (such as ChatGPT and other open-ended \acp{LLM}) can give learners more flexibility and engage them to direct their own learning.
However, they cannot guide learners towards larger learning goals and might provide information that is inaccurate.
On the other end of the spectrum, completely scripted dialog systems (such as pre-written linear dialogs~\cite{friedman1995little,friedman1995seasoned}) can ensure that learners meet pre-determined learning goals and give accurate information.
However, they can also limit learner agency and creativity.

\platform{} aims to strike a balance.
It uses pre-written dialog trees (which can be guaranteed to meet pre-determined learning goals with accurate information) while allowing open-ended \ac{LLM} queries.
In essence, the dialog trees act as ``rails'' to keep learners on track, but students can go ``off-road'' by asking their own questions.
These ``rails'' can also help readers ensure that they are on the right path to achieve the target learning goals.
Providing both dialogic structure and agency represents a novel approach to aligning open-ended \ac{LLM} interactions with pedagogical goals.
Future advances in \ac{LLM} accuracy, alignment, and goal planning might enable less constrained systems in the future.

\subsubsection{Dialog Representation}
The next fundamental question is how to represent dialogs while ensuring learners meet pre-determined learning goals.
An essential criterion is that \emph{any information related to the learning goals must be accurate}.
Despite their capabilities, \acp{LLM} can still generate inaccurate information.
Thus, any content related to the core learning goals should be either written or verified by \platform{} authors.
This is especially important for assessments (e.g., coding and multiple-choice questions), where we anecdotally found that \acp{LLM} often wrote assessments that were \emph{accurate} but missed the primary learning goals.

Because \platform{} authors are responsible for verifying or writing learning material, dialog representations with large state spaces (e.g., state-based representations) or that are not predictable (e.g., using topic-shifting or open-ended dialogs) would not be appropriate.
We instead designed \platforms{} to represent dialogs as trees, with a single `target' leaf node that represents completion of the learning goals.
Each node in the tree represents a dialog message, where the root node is the starting prompt, and child nodes are possible responses.
This representation of dialogs reduces the amount of information that authors need to verify (only the nodes from the `root' to the `target').
Using dialog as trees rather than linear scripts also supports dynamic narratives with user agency.

\begin{figure}
    \centering
    \includegraphics{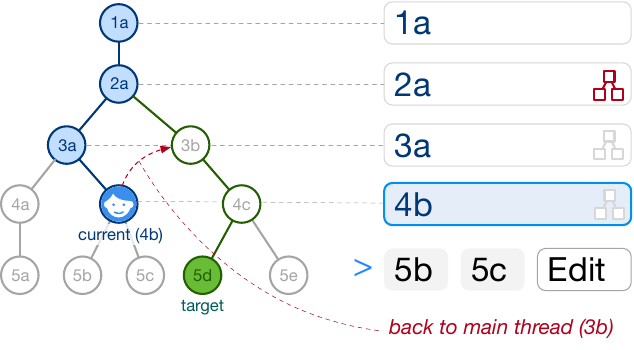}
    \caption{Conceptually, \platform{} dialogs are trees (left) but these trees are visually simplified and flattened in the reader \ac{UI} (right) for readability.
    On the left is an illustration of an example dialog tree.
    Each node in the tree represents a message and is labeled according to its depth (1--5) and breadth-first search order (a--z).
    For example, 3a represents the leftmost node in the third layer.
    \platform{} allows instructors to define a `target' node ({\color{oggreen}message 5d} in this example) that represents completion of the learning goals for the page.
    In this example, the user is currently focused on {\color{ogblue}message 4b}, so `ancestor' cells {\color{ogblue}1a}, {\color{ogblue}2a}, {\color{ogblue}3a}, and {\color{ogblue}4b} are visible in the \ac{UI} (right).
    In the \ac{UI}, users can also select pre-existing subsequent messages (5b and 5c) or write a custom message (`Edit') that will query an \ac{LLM} to add a response message.
    The user's current node is part of the dialog tree but not on the path of the target, {\color{oggreen}5d}.
    Thus, the user is also nudged to go to a node that would move them back towards {\color{oggreen}target node 5d} by back-tracking to {\color{ogred}node 3b} (with a {\color{ogred}``back to main thread'' button}).
    The \ac{UI} also shows icons (\branch) to indicate messages with multiple possible responses (right) and calls attention to the first message where the user's path diverges away from the target message (in {\color{ogred}red \branch{}}).}
    \label{fig:dialog-tree}
\end{figure}

For readers, these dialog trees are simplified and flattened to ensure the dialog is coherent and readable.
As Figure~\ref{fig:dialog-tree} illustrates, readers can see their current node and all its ancestors, along with icons to indicate where there are multiple options to explore.
Readers can explore tangents before being guided back to the target learning path.
As readers progress, the \ac{UI} highlights divergence from this target path, with a link to go ``\emph{back to the main thread}''---to the first node that branched away from the target node.


\subsubsection{Using and Managing \ac{LLM}-Generated Content}
A key design feature of \platform{} is in the way that it integrates \acp{LLM} with pre-written narratives.
Both authors and readers can use \acp{LLM} in \platform{}.
%
Authors can use \acp{LLM} to generate an initial dialog tree.
They first specify a list of participants (by default, one `instructor' persona, named \emph{Dr. Ed} and a persona representing the reader).
Each of these personas includes a description, which can be supplied to the \ac{LLM} to generate appropriate dialog.
Most importantly, the description of the instructor persona should be aligned with the desired learning goals---for example, ``A knowledgeable instructor teaching the basics of \texttt{for} loops in Python in a way that is suitable for novices.''.
Authors can then either generate dialogic content turn-by-turn---for example, by adding messages on behalf of the `student' persona and letting ChatGPT respond as the `instructor' persona.
Authors can also leverage ChatGPT to generate multi-turn dialogs, by describing the dialog topic and the participants' personas.
In both cases, authors can manually edit the \ac{LLM}-generated content as necessary.
In our experimentation, we found mixed results with fully \ac{LLM}-generated content.
When asked about popular topics, it generated content that was largely accurate but the generated messages were not always engaging for readers (for example, without further guidance, messages would often be longer than what we found most readers would prefer) and the generated assessments were not always aligned with the learning goals (for example, there might be multiple choice questions focused on terminology rather than core conceptual ideas).

\begin{wrapfigure}{r}{0.5\textwidth}
    \centering
    \includegraphics[width=0.5\textwidth]{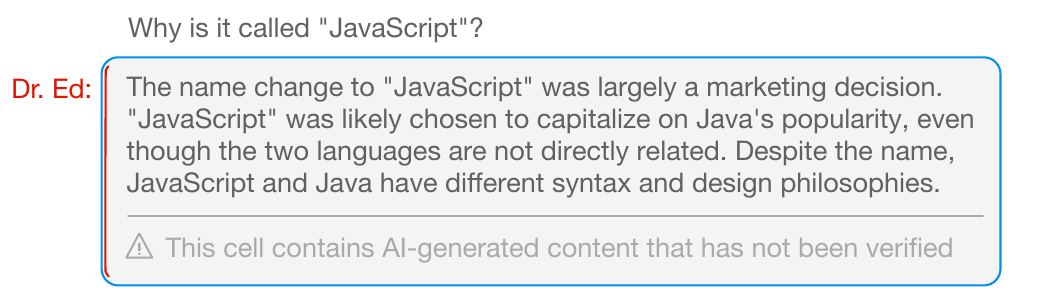}
    \caption{Warning labels are added to any cells that contain un-verified \ac{LLM}-generated content.}
    \label{fig:ai_content_warning}
\end{wrapfigure}

\platform{} readers can engage with \acp{LLM} at any point in the dialog, as necessary.
Whenever they query an \ac{LLM}, the current node and all ancestor nodes are passed in as context.
The \ac{LLM}'s response then creates a new branch in the dialog tree.
By default, the \ac{LLM}'s response is associated with the `instructor' persona (Dr. Ed) but any \ac{LLM}-generated content is also accompanied by a warning that the content is \acs{AI}-generated and has not been verified, as Figure~\ref{fig:ai_content_warning} illustrates.

\subsection{Complementary Features}\label{ssec:complementary}
\platform{} also contains many features that complement its core interactions.
These features are not novel individually but are crucial for the effectiveness of \platform{} as a learning tool.

\subsubsection{Incremental Code Changes}
Many instruction materials contain larger code samples as part of the learning material.
Because these code samples are meant to be instructive, they should be constructed in a way that makes them as understandable as possible.
As prior work has found, tutorial authors often want to incrementally build didactic code examples and explain these code samples in an order that is different from the ``code order''~\cite{head2020composing}.

\begin{wrapfigure}{r}{0.5\textwidth}
    \centering
    \vspace{-10pt}
    \includegraphics[width=0.5\textwidth]{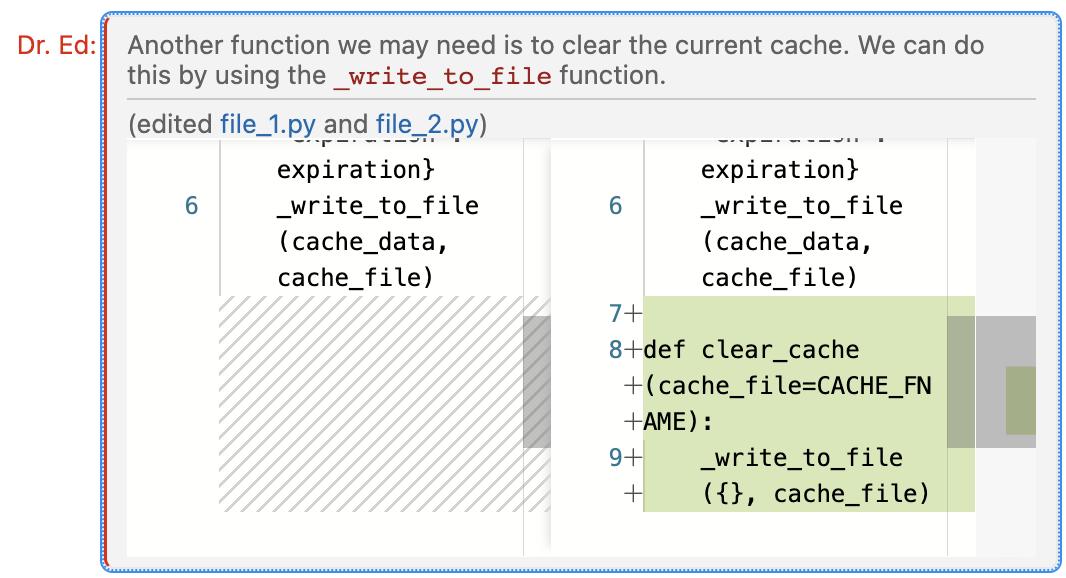}
    \caption{Instructors can work with larger codebases that change over time.
    Readers can see these incremental code edits, as diffs.}
    \label{fig:diff}
\end{wrapfigure}

To support this, \platform{} allows authors to build up larger code examples incrementally over the course of a narrative.
Each message can be associated with a textual code diff representing additions, removals, or unmodified lines relative to prior code versions.
These diffs are rendered to visually show the code evolving across messages.
This allows authors to introduce concepts through iterative code refinements.
\platform{} also displays code changes inline with messages, as Figure~\ref{fig:diff} illustrates.

\subsubsection{Deictic References}
It is important to be able to orient readers to understand code samples as they are being incrementally built.
But it can be difficult for readers to understand which message refers to which part of code, particularly for larger code samples.
Prior work~\cite{wang2020callisto,oney2018creating} addresses this through `deictic' references---visual indicators that associate messages with the part of code it is discussing.
Similarly, EDBook allows authors to create deictic pointers, which are displayed as curved lines between messages and regions of code, as Figure~\ref{fig:edbook-main} illustrates.

\subsubsection{Interactive Assessments}
To check and reinforce reader comprehension, authors can incorporate interactive assessments into \platform{} narratives.
Currently, \platform{} supports multiple-choice questions and code writing challenges.
These assessments provide opportunities for active learning and feedback.
Learners can validate their knowledge through practice.
Incorrect answers reveal areas needing more focus.

\subsection{Novelty and Uniqueness}
As far as we are aware, \platform{} is the first platform to combine pre-written dialog trees with context-informed \acp{LLM}.
This design mitigates unhelpful \ac{LLM} responses and keeps learners focused while allowing flexibility.
The dialog trees provide goal-oriented structure that we can guarantee to be accurate while the \ac{LLM} queries enable personalized explorations.
The technique is novel in using dialog state for education-oriented \ac{LLM} guidance.
Further, the design of \platform{}'s deictic references (visually tying messages with the code being discussed) is unique relative to other tools that contain deictic code references~\cite{wang2020callisto, oney2018creating}.

Finally, complementary capabilities (incremental code edits, deictic references, and interactive assessments) enrich the core dialogic learning in \platform{}.
They demonstrate how interactive features can be effectively integrated to engage learners.
The combination enables more pedagogically productive programming narratives.
\section{User Study}To evaluate the effectiveness of \platform{} and identify usage patterns, we conducted a within-subjects user study with 20 programming learners.
Our study focused on the \emph{reading} experiences of learners, as opposed to the authoring experience.
We view the reading experience as disproportionately important because we expect many more users to be readers than authors.


\subsection{Recruitment}
We reached out to students enrolled at \anonymized{(INSTITUTIONS REMOVED FOR ANONYMITY)} through email.
Participants were required to have taken at least one Python programming course but no prior experience with the specific content covered in the tutorials---no experience with Scheme or LISP (one task in our study) and no experience with writing code for caching web requests in Python (another task in our study).
We determined eligibility by participants' responses on a screening form.

\subsection{Participants}

Of the eligible participants, 20 completed the study.
We selected participants based on their eligibility and aimed for a diverse set of backgrounds.
One additional participant began the study but stopped the study after performing one task due to internet connectivity problems.
All participants were students, as learners are the primary audience for \platforms{}.
Most participants (16 of 20) were in degree programs that involved programming (either Computer Science or Information), as we required some prior technical knowledge.
Relevant demographics of the 20 remaining participants are itemized below:

\begin{table}[htbp]
\centering
\begin{tabular}{
  c
  >{\centering\arraybackslash}p{2cm}
  c
  >{\centering\arraybackslash}p{3cm}
  >{\centering\arraybackslash}p{3cm}
  >{\centering\arraybackslash}p{2.5cm}
}
\toprule
\textbf{Participant \#} & \textbf{Age} & \textbf{Gender} & \textbf{Education Program} & \textbf{Prog Experience} & \textbf{LLM Usage} \\
\midrule
1 &18--24 &Woman &Undergraduate &Less than one year &Rarely \\
2 &18--24 &Woman &Master's &1--2 years &Occasionally \\
3 &18--24 &Man &Undergraduate &1--2 years &Occasionally \\
4 &18--24 &Woman &Undergraduate &1--2 years &Occasionally \\
5 &25--34 &Woman &Ph.D. &3--5 years &Frequently \\
6 &18--24 &Man &Undergraduate &Less than one year &Rarely \\
7 &18--24 &Woman &Master's &Less than one year &Occasionally \\
8 &18--24 &Woman &Undergraduate &1--2 years &Frequently \\
9 &25--34 &Woman &Master's &Less than one year &Occasionally \\
10 &18--24 &Woman &Undergraduate &6--10 years &Rarely \\
12 &25--34 &Woman &Master's &3--5 years &Frequently \\
13 &18--24 &Woman &Master's &3--5 years &Rarely \\
14 &25--34 &Man &Master's &6--10 years &Frequently \\
15 &18--24 &Man &Master's &1--2 years &Frequently \\
16 &18--24 &Man &Undergraduate &1--2 years &Rarely \\
17 &25--34 &Man &Ph.D. &3--5 years &Occasionally \\
18 &25--34 &Man &Ph.D. &1--2 years &Occasionally \\
19 &18--24 &Woman &Master's &Less than one year &Frequently \\
20 &18--24 &Woman &Undergraduate &1--2 years &Occasionally \\
21 &18--24 &Woman &Ph.D. &1--2 years &Rarely \\
\bottomrule
\end{tabular}
\caption{Participants' Data. Participant 11 did not complete the study and is thus not included in this table or our analysis.}
\end{table}

All participants had prior experience using ChatGPT and 15 of 20 had used ChatGPT in their own learning, primarily for programming support.
All but one participant had used Visual Studio Code.

\begin{figure}
    \centering
    \includegraphics[width=0.8\textwidth]{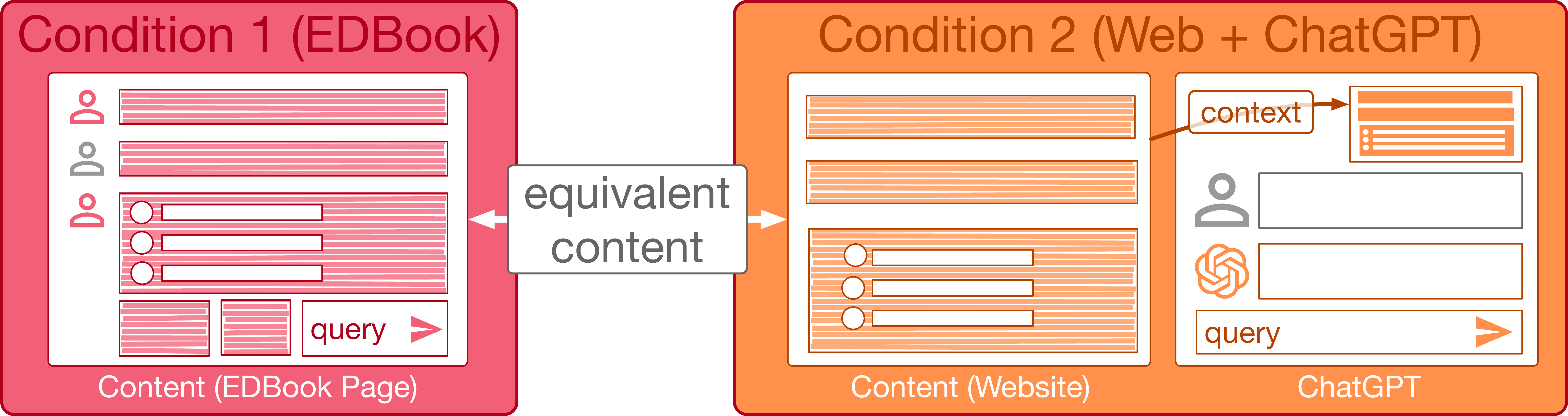}
    \caption{An Illustration of the two conditions in our evaluation of \platform{}.
    In \condedbook{} (left), participants performed a task with the \platform{} platform.
    In \condbaseline{} (right), participants performed a task with equivalent content, converted to a webpage.
    \condbaseline{} participants can also interact with a ``context-informed'' instance of ChatGPT that has the context of the relevant webpage content to provide contextually-appropriate responses.
    }
    \label{fig:study-conditions}
\end{figure}

\subsection{Setup and Procedure}
Our study had two conditions, illustrated in Figure~\ref{fig:study-conditions}:

\begin{itemize}
    \item Condition 1 (\condedbook{}): Participants read educational dialogic \platform{} content.
    \item Condition 2 (\condbaseline{}): Participants read the same educational content but in non-dialogic format and translated into a standard webpage format. They also had access to ChatGPT, with the content provided as background context to ensure it produced responses that were consistent with the content.
\end{itemize}

And two tasks:

\begin{itemize}
    \item Task 1 (\taskscheme{}): In this task, participants learned the basics of the Scheme programming language (a functional language that is a dialect of Lisp).
    This task was designed to represent the challenges of \emph{learning content that is conceptually new}, as most participants have no prior familiarity with any form of declarative programming.
    \item Task 2 (\taskcache{}): In this task, participants walked through how to write Python code that makes web requests (using the \texttt{requests} module) and caches the result for a pre-specified period of time.
    Participants had familiarity with the underlying programming language (Python) but not with this specific application or algorithm.
    This task was designed to represent the challenges of \emph{learning through building}. Although none of the underlying concepts were new to participants, the specific way they were put together in a larger (approximately 70-line) code sample could be difficult to follow.
\end{itemize}

In both conditions, tasks were split into smaller topic-focused chapters.
These chapters included quiz questions and other methods to allow readers to check their understanding.
These quiz questions were different from the assessment questions we used after each task to asses their understanding.
In both conditions, participants interacted with OpenAI's ChatGPT-3.5.

\begin{figure}
    \centering
    \includegraphics[width=\textwidth]{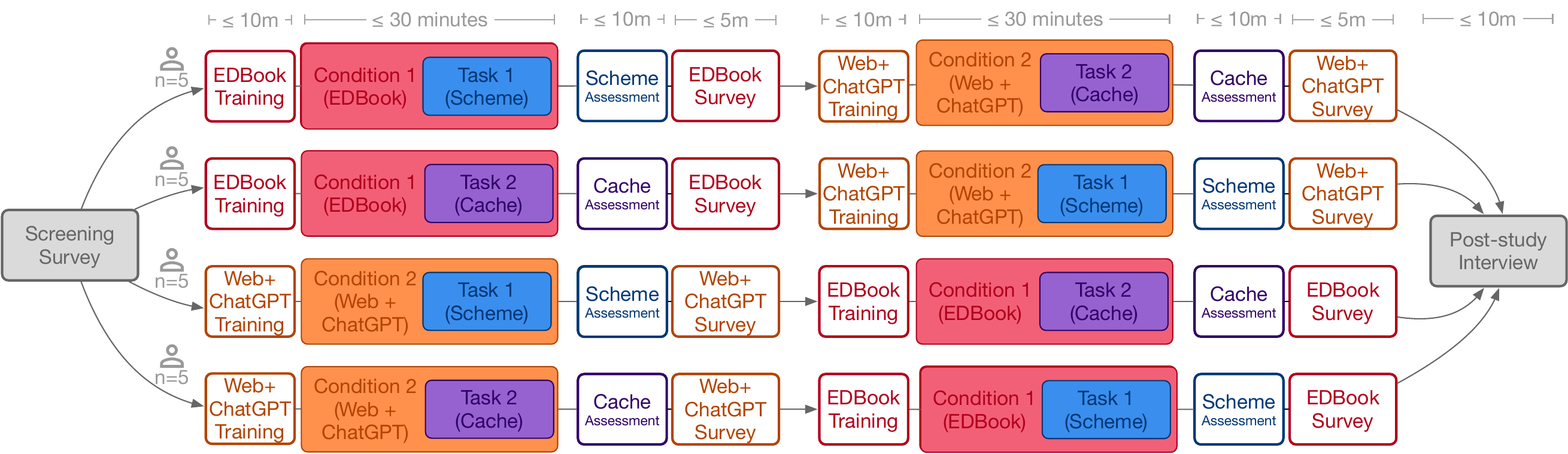}
    \caption{An Illustration of the steps that participants take in our study. Each participant performs two tasks using two different conditions. They complete post-task surveys and interviews.}
    \label{fig:study-setup}
\end{figure}

Figure~\ref{fig:study-setup} illustrates our study design.
The study coordinator began each study by asking participants to complete a consent form and informing them of the study procedure.
They then completed two learning tasks (\taskscheme{} and \taskcache{}) in both conditions (\condedbook{} and \condbaseline{}), in a latin square design to counter order effects.
Before each task, participants spent approximately 10 minutes walking through the interface(s) they would interact with for that condition (instructions on how to use \platforms{} or ChatGPT).
After each task, participants completed assessments to gauge their understanding of the materials.
They were given 10 minutes for each assessment and any incomplete answers were marked as ``incorrect''.
During these assessments, they were allowed to go back and check the learning materials but they were not allowed to send further queries to the \ac{LLM} (to prevent participants from feeding assessment questions back to the \ac{LLM} without demonstrating an understanding of the material).
We also asked participants to fill out a short survey after each task.
After participants completed both tasks, the study coordinator conducted a post-task interview.

All study sessions were conducted virtually.
With participants' approval, the study coordinators recorded their screens and collected other usage telemetry information for subsequent analyses.
Each session took approximately two hours and participants received \$50 (USD) as compensation.
Our institution's \ac{IRB} approved our study setup.

%

\subsection{Results}
We describe our quantitative results below and summarize our interview responses and findings in section~\ref{ssec:interview_and_findings}.

\subsubsection{Measures of Learning Activities}
As Table~\ref{tab:quant_measures} shows, we collected several measures of the learning activities, including the number of questions participants asked with \acp{LLM}, the time they spent on learning, the time they spent on assessments, assessment scores, and the number of attempts to try example code.
We found that participants in the \condedbook{} spent significantly longer time with the learning materials ($\overline{x}$ = 1817.85, $\sigma$=265.22) than in the \condbaseline{} ($\overline{x}$ = 1389.90, $\sigma$=461.18).
In addition, participants in the \condedbook{} ($\overline{x}$ = 0.95, $\sigma$=0.13) made more attempts to answer the quiz questions and experiment with the example code than those in the \condbaseline{} ($\overline{x}$ = 0.40, $\sigma$=0.36).

\subsubsection{Types of Questions Asked with \acp{LLM}}
As Table~\ref{tab:question_types} shows, we manually coded the questions that participants asked with \acp{LLM}.
We used Bloom's Taxonomy~\cite{bloom1956bloom} for closed coding, including six levels---remember, understand, apply, analyze, evaluate, and create.
In both conditions, we did not observe questions related to remembering (recalling facts and basic concepts).
In \condedbook{}, participants asked more questions for seeking usage of a concept in new situations (applying) and differentiating between two concepts (analyzing), whereas in \condbaseline{}, participants asked more questions for seeking explanations of a code solution (evaluating) or directly requesting a solution for a coding problem (creating).
This could indicate that \platforms{} led participants to actively reflect on the concepts they learned.
This reflection includes considering edge cases, evaluating generalizability, and connecting with previously learned concepts.
In contrast, in \condbaseline{}, participants adopted a more passive learning approach, primarily seeking solutions to quiz questions from \acp{LLM} and seldom experimenting with these solutions within the provided \ac{IDE}.

\subsubsection{Post-task Questionnaire Results}
As Table~\ref{tab:survey_result} shows, we examined the post-task questionnaire results.
Among all the measures, we found that in \condedbook{}, participants gave a significantly higher rating to the question \say{I was highly engaged in the tutorial when learning} ($p<0.05$, ANOVA).
Participants also reported that the code-writing questions are more helpful in \condedbook{} than \condbaseline{} ($p<0.05$, ANOVA).

\begin{table}[t]
\newcommand{\vislen}{5cm}
\begin{tabular}{ccccp{5cm}} 
 \toprule
 \textbf{Measures}&\textbf{Condition} & \textbf{$\overline{x}$} & \textbf{$\sigma$} & \\
 \midrule
  \multirow{2}{*}{Number of Questions Asked with LLMs} & 
  \condbaseline{} & 3.45 & 3.89 & \multirow{2}{*}{\hspace{-10pt} \includegraphics[width=\vislen]{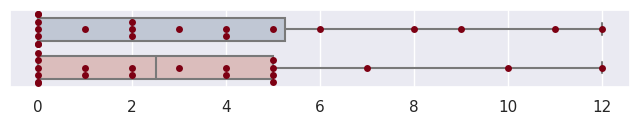}} \\
 &\condedbook{} & 3.30 & 3.45 & \vspace{5pt} \\
  \midrule
  \multirow{2}{*}{Time Spent on Learning* ($p<0.01$)} &  \condbaseline{} & 1389.90 & 461.18 
  & \multirow{2}{*}{\hspace{-13pt} \includegraphics[width=\vislen]{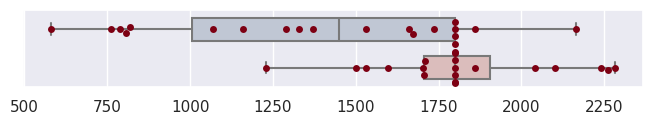}}\\
 & \condedbook{} & 1817.85	 & 265.22  &\vspace{5pt} \\
 \midrule
 \multirow{2}{*}{Time Spent on Assessment} & \condbaseline{} & 575.0 & 39.38  & \multirow{2}{*}{\hspace{-10pt} \includegraphics[width=\vislen]{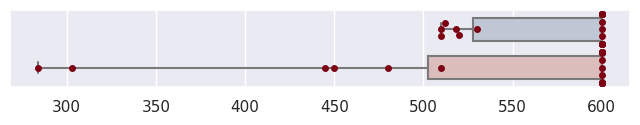}} \\
 & \condedbook{} & 543.6 & 100.91 & \vspace{5pt}\\
 \midrule
 \multirow{2}{*}{Assessment Score} & \condbaseline{} & 1.35 & 0.74 &   \multirow{2}{*}{\hspace{-10pt} \includegraphics[width=\vislen]{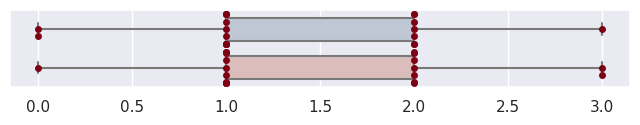}} \\
 & \condedbook{} & 1.50 & 0.76  &\vspace{5pt} \\
 \midrule
 \multirow{2}{*}{Attempts to Try Example Code* ($p<0.001$)} &  \condbaseline{} & 0.40 & 0.36  &   \multirow{2}{*}{\hspace{-10pt} \includegraphics[width=\vislen]{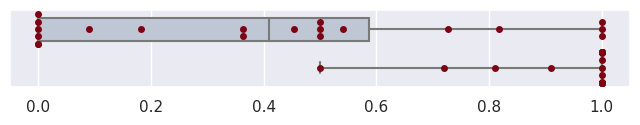}}\\
 &\condedbook{} & 0.95 & 0.13& \vspace{5pt}\\
 \bottomrule
\end{tabular}
\caption{We collected the number of questions participants asked with LLMs, the time they have spent on learning and the exit test, the test score, and the number of attempts to try quiz questions. (mean: $\overline{x}$, standard deviation: $\sigma{}$). 
Our comparison suggests that the time spent on learning is significantly different in the two conditions ($p < 0.01$, ANOVA); the number of attempts to try quiz questions is also significantly different in the two conditions ($p < 0.001$, ANOVA). 
}
\label{tab:quant_measures}
\end{table}

\begin{table}[t]
    \centering
    \begin{tabular}{cm{4.6cm}m{5.2cm}c}
     \toprule
      \multirow{2}{1.3cm}{\centering\textbf{Code}} & 
      \multirow{2}{4.6cm}{\centering\textbf{Definition}} & 
      \multirow{2}{5.2cm}{\centering\textbf{Example}} & \condbaseline{} vs. \condedbook{} \\ & & & \includegraphics[width=2.88cm]{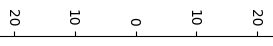}\\
    \midrule
       Understand  & Seeking explanation of a concept instead of a real coding problem  & ``What is private function in Python?'' & \multirow{6}{*}{\includegraphics[width=2.88cm]{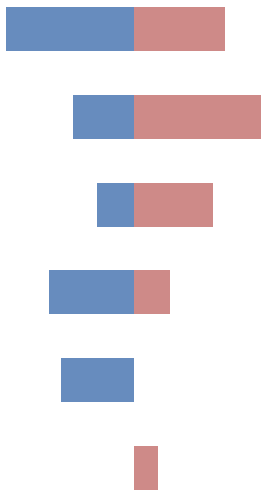}} \\
       \midrule
        Apply & Seeking usage of a concept in new situation & ``Can you perform multiple operations in one s-expression?'' \\
        \midrule
        Analyze & Differentiating between two concepts & ``Is \texttt{write} or \texttt{display} more similar to a \texttt{print} function in Python?'' \\
        \midrule
        Evaluate & Explaining a coding solution (e.g., asking explanations of the answers) & ``Why are there a lot of \texttt{if} statements in the code?'' \\
        \midrule
        Create & Directly requesting a solution for a coding problem & \textit{(copy and pasted code writing question to get an answer)} \\
        \midrule
        Others & Conversations instead of questions & ``I've learned the concept of caching before, but not familiar with detailed info'' \\
        \bottomrule
    \end{tabular}
    \caption{Types of Questions Participants Asked to LLMs}
    \label{tab:question_types}
\end{table}

\subsection{Interview Responses and Findings}\label{ssec:interview_and_findings}

Based on the results described above and interviews with participants, we identified several key findings.

\subsubsection{Participants used more time for learning the materials in \condedbook{} than \condbaseline{}}\label{sssec:more_time}

As our results show, participants spent more time in \condedbook{}  than \condbaseline{} on average ($p < 0.01$).
Based on our observations and intuition, we believe there are several reasons for this.
First, the \platform{} platform requires at least some user interaction to reveal content, whereas \condbaseline{} requires no interaction beyond scrolling.
Second (and possibly consequentially), participants attempted more coding and multiple-choice exercises in \condedbook{} than in \condbaseline{}, as we describe in \ref{sssec:more_quizzes}.
As we found in interviews, participants liked the ability to test out code immediately in \condedbook{}, which might have led to increased engagement and more time spent on coding quizzes.
Third, as section \ref{sssec:more_engaging} describes, participants expressed that they felt more engaged in \condedbook{} than in \condbaseline{}.
This increased engagement might have led to more time spent in \condedbook{}.
Finally, \platforms{} were new to all participants and it takes some time to get acquainted with any unfamiliar tool, despite having a tutorial.

\subsubsection{Participants attempted more coding quizzes in \condedbook{} than \condbaseline{}}\label{sssec:more_quizzes}

Our results also show that participants attempted more coding quizzes in \condedbook{} than \condbaseline{} ($p < 0.001$).
Again, this might be because \platforms{} require user interaction to advance the narrative.
This means that participants needed to engage with coding question at some level (the need to explicitly decide to ``skip'' the quiz or reveal the answer).
It might also be because the \platform{} platform integrates tutorial content with a code authoring and testing environment:

\participantquote{[With \platforms{},] I could write and test [code] in the same page. [In \condbaseline{}, I] had to switch between code runner and the content}{P12}

Participants also expressed in surveys that they found code writing questions in \condedbook{} to be more useful than in \condbaseline{}, with significantly higher agreement with the survey question ``The code writing questions helped me improve my understanding of the material'' ($p<0.05$).
This might also be due to \platforms{} giving richer feedback to code writing responses.
In \condbaseline{}, several participants sought richer feedback by copying and pasting quiz questions into ChatGPT rather than scrolling to see the answer.

\subsubsection{Participants found it easier to phrase/ask questions in \condedbook{} than \condbaseline{}}\label{sssec:easier_to_ask}

Although ChatGPT in \condbaseline{} was provided the context of the content being discussed, this was a \emph{global} context.
However, participants felt that in \condbaseline{}, it could be burdensome to provide the \emph{local} context of their queries.
Further, combining \ac{LLM} input with content made P13 more confident that the \ac{LLM} was prompted with the appropriate context:

\participantquote{I know you said like in this one (\condbaseline{}) you had prompted it with all the information that it would need, but it was harder. [...] like I wanted to ask a specific question about this function, and it kind of knew which function I meant. But it was like assuming the function, so I didn't know if it knew what I was referencing. So in this one (\platform{}) I was a lot more confident that I was like: in this example, what do you mean by this thing? And there’s more one-to-one correlation of the question versus the information I'm saying.}{P13}

This local context could be crucial for understanding how to accurately respond to some of the brief questions that participants asked during our study, such as \inlineparticipantquote{what is url}{P20}, \inlineparticipantquote{Can you give an example of it}{P15}, or \inlineparticipantquote{what does cube mean}{P12}.
For some of these queries, ChatGPT might give accurate answers but without being confident about the local context, it can also respond with superfluous information.
For example, ``what is url'' refers to a variable \texttt{url} in a code sample but without local context, ChatGPT respond to ``what is url'' by first describing what \acsp{URL} are.
One participant also expressed that they would feel more comfortable using \acp{LLM} through \platforms{} in a classroom setting:

\participantquote{I actually stay away from using like ChatGPT, because I feel like so much debate about it, and so many like cheating things related to it. [...] I feel like in an unregulated ChatGPT if I'm the one putting all the input in and not getting as personalized output, it feels a little bit risky like am I crossing the line? But with this (\platform{}), like everything was prepared for me, [...] it felt a lot safer if was an assignment or something. I wouldn't feel guilty about using it because it was created for me. [...] I really like the academic safety of it.}{P10}

\begin{table}[]
    \centering
    \footnotesize
    \begin{tabular}{p{4.5cm}cc}
        \toprule
         \textbf{Statement} & \textbf{Condition} & \includegraphics[width={9.5cm}]{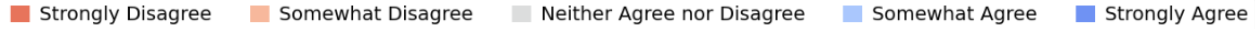} \\
         \midrule
         \multirow{2}{=}{I understand all the concepts and codes in the tutorial.} & \condbaseline{} & \multirow{2}{*}{\includegraphics[width={9.5cm}]{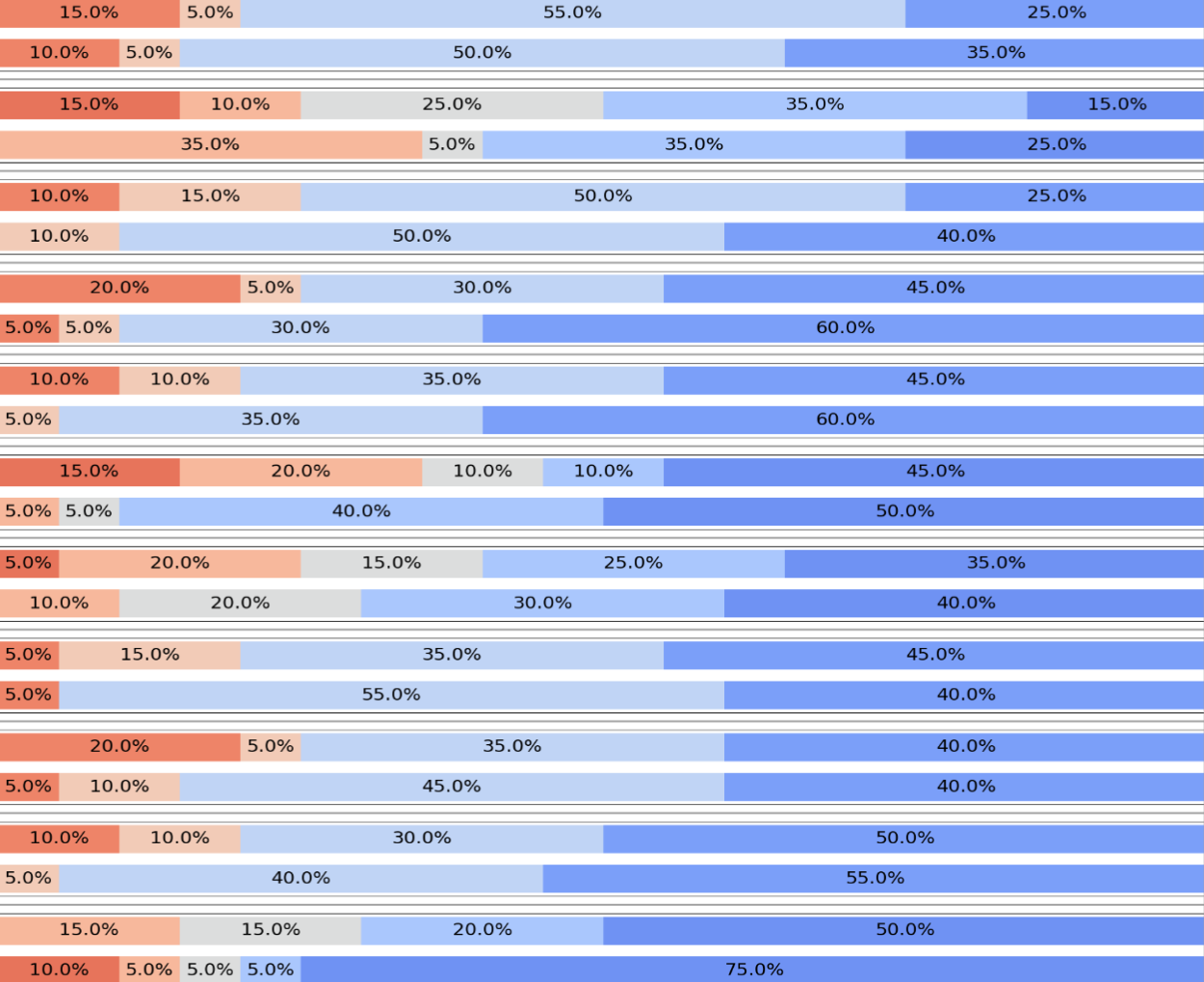}}\\
         & \condedbook{} & \\
         \midrule
         \multirow{2}{=}{I am confident in writing functions similar to what I learned in the tutorial.} & \condbaseline{} & \\
         & \condedbook{} & \\
         \midrule
         \multirow{2}{=}{The tutorial explained the concept clearly without any confusion.} & \condbaseline{} & \\
         & \condedbook{} & \\
         \midrule
         \multirow{2}{=}{The tutorial is easy to follow and understand.} & \condbaseline{} & \\
         & \condedbook{} & \\
         \midrule
         \multirow{2}{=}{The tutorial is well structured with a suitable length.} & \condbaseline{} & \\
         & \condedbook{} & \\
         \midrule
         \multirow{2}{=}{I was highly engaged in the tutorial when learning.} & \condbaseline{} & \\
         & \condedbook{} & \\
         \midrule
         \multirow{2}{=}{The tutorial met my learning preferences and needs.} & \condbaseline{} & \\
         & \condedbook{} & \\
         \midrule
        \multirow{2}{=}{I found it easy to navigate through the content of the tutorial.} & \condbaseline{} & \\
        & \condedbook{} & \\
        \midrule
        \multirow{2}{=}{The code examples were easy to follow and understand.} & \condbaseline{} & \\
        & \condedbook{} & \\
        \midrule
        \multirow{2}{=}{The multiple-choice questions helped me improve my understanding of the material.} & \condbaseline{} & \\
        & \condedbook{} & \\
        \midrule
        \multirow{2}{=}{The code-writing questions helped me improve my understanding of the material.} & \condbaseline{} & \\
        & \condedbook{} & \\
        \midrule
        \multirow{2}{=}{Code pointers (which highlighted bits of code) were useful.} & \condedbook{} & \multirow{2}{*}{\includegraphics[width={9.5cm},height={2cm}]{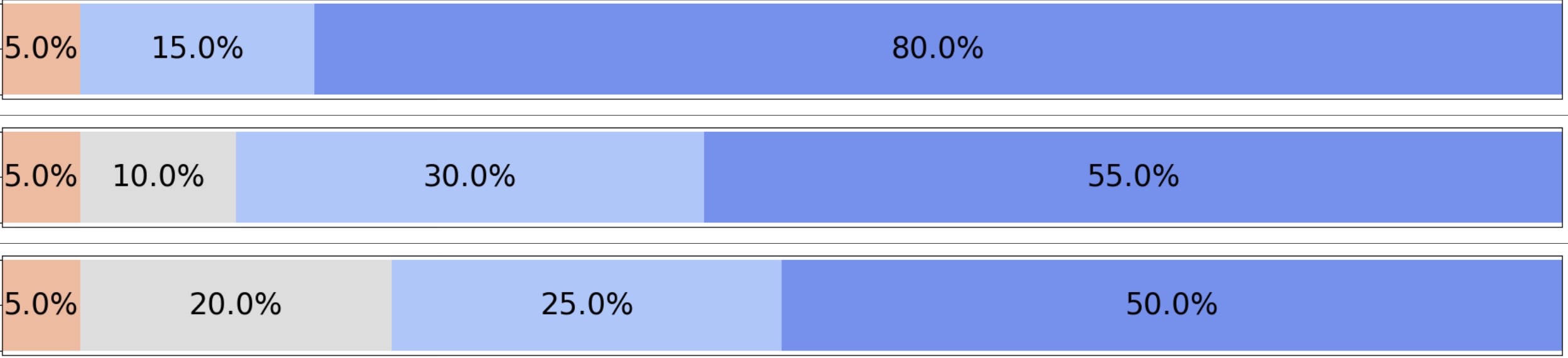}}\\
        & & \\
        \midrule
        \multirow{2}{=}{The ability to select between multiple possible messages was useful.} & \condedbook{} & \\
        & & \\
        \midrule
        \multirow{2}{=}{The ability to write custom messages is useful.} & \condedbook{} & \\
        & & \\
        \bottomrule
    \end{tabular}
    \caption{Post-task questionnaire results for participants in both conditions.}
    \label{tab:survey_result}
\end{table}

\subsubsection{The types of questions that participants asked in \condedbook{} were different from those in \condbaseline{}}

We found qualitative differences in the types of questions that participants asked in the two conditions.
In \condbaseline{}, we observed more participants copying and pasting a local context from the content (\ref{sssec:easier_to_ask}).
Some participants in \condbaseline{} also copied and pasted quiz questions to get more feedback (\ref{sssec:more_quizzes}).
We found that \ac{LLM} queries in \condedbook{} tended to focus more on applying conceptual knowledge and exploring edge cases of concepts.
We also observed two participants in \condedbook{} who used the \ac{LLM} to have conversations that were more tangential to the learning content---for example, to describe their background.

\subsubsection{Participants found it easier to navigate and search the contents in \condbaseline{}}

Participants expressed that they found it easier to navigate and search tutorial content in \condbaseline{}.
There were two primary reasons for this.
First, our \platform{} content was written as a dialog and thus contained some messages that advanced the dialog but were tangential to the learning goals.
Some participants felt that this was superfluous:

\participantquote{Some of the suggestions felt like a little bit fluffy. [...] I think they mostly care about learning language and just getting to it. [...] I think it could have been reduced just to get rid of those like fluffy questions, I would just rather get to the point.}{P10}

Second, the ``find on page'' and skimming interactions in the \platform{} platform are not as refined as they are in commercial web browsers.
This could make navigation more difficult for some participants:

\participantquote{Whenever I have questions [in \condbaseline{}] I can just do a \texttt{Ctrl+F} and go back to the previous content. So it's more free and it's easier to navigate for me.}{P9}

Despite this interview feedback from some participants, there were not significant differences in responses to survey questions about the ease of navigation in \condedbook{} compared to \condbaseline{}.

\subsubsection{Participants found \condedbook{} to be more engaging}\label{sssec:more_engaging}

When asked to compare \condedbook{} with \condbaseline{}, most (13) participants independently expressed that they found \platforms{} to be more engaging that ChatGPT (our interview question did not directly ask about engagement).
Some representative quotes from participants:

\participantquote{So I'm a lot more engaged for [\condedbook{}], but [in \condbaseline{}] I kind of just lost track of what I was reading.}{P1}


\participantquote{I think I kind of prefer [\condedbook{}] learning experience. I think, like more engaging and interactive.}{P4}

\participantquote{The part that I like about [\condedbook{}] is that the interactive element is helping me keep me stay engaged with the materials.}{P5}

\participantquote{I like the dialogue style, because I feel like, I mean, obviously, I'm aware that it's not a real person, but I think it's I like the way that the instructor is talking, and it's kind of more engaging than just like reading the text. }{P13}




This increased engagement is also supported by our survey results, where participants' self-rated agreement with the statement ``I was highly engaged in the tutorial.'' was significantly higher in \condedbook{} than \condbaseline{} (p<0.05).
However, two participants disagreed with this sentiment, expressing that pre-written responses discouraged them from engaging in critical thinking:

\participantquote{I become lazy---I know it'll always give me a suggested message.}{P9}


\subsubsection{Participants would prefer \platform{} content when learning something for the first time and a standard \ac{LLM} \ac{UI} when they were familiar with the material.}\label{sssec:prefer_when_first_time}
In post-task interviews, we asked participants to speculate on when they would prefer \platform{} content and when they would prefer `standard' content and \acp{LLM}.
The most common answer (from nine participants) is that they would prefer \platforms{} when learning material for the first time and `standard' \ac{LLM} interactions when they are already familiar with the material:

\participantquote{[\platforms{}] will be helpful when I'm learning a language for the first time but [ChatGPT] will be better when I already have prior knowledge in it}{P2}

\participantquote{If I were a new student like my first year of programming, if I had something like [\platforms{}], I would really like it. Because learning it the traditional way was quite difficult for someone who's starting, but if I had this on day one, I would use it a lot.}{P15}

This feedback is also supported by the survey results, where participants responded that \platform{} content was easy to follow and found the quiz questions more useful in \condedbook{}.
For \taskcache{}, \condedbook{} had significantly higher ratings for both questions than \condbaseline{}.
It is also supported by related interview responses, where participants expressed that \platforms{} were more engaging (\ref{sssec:more_engaging}).

\subsection{User Evaluation --- Limitations and Discussion}

Our results provide important insight into how participants interact with educational content (both \platforms{} and ``standard'' content with ChatGPT).
However, as with any controlled study, there are important limitations to consider with respect to the results.
First, our participant pool is not representative of the demographics of all learners.
Second, participant feedback to \platform{} content might have been more positive for participants who wanted to play the role of a ``good participant''.
Finally, our study was a short term study and is not necessarily representative of longer-term usage patterns.

Still, we believe our results are indicative of realistic interaction patterns.
Overall, we found that participants appreciated many of the interactive parts of \platforms{} and found them more engaging.
In our study, this increased engagement did not lead to improved assessment scores (assessment score averages in \condedbook{} were higher and time spent on assessment in \condedbook{} was lower but neither difference with \condbaseline{} was statistically significant).
However, many prior studies have found that increased learner engagement improves learning outcomes~\cite{gray2016effects}.

Participants also pointed to ways that the \platform{} platform could improve: (1) improving navigability and (2) promoting critical thinking.
We believe that navigability (1) could be improved with relatively small implementation changes---writing more concise dialogic content and engineering better ``search'' features.
However, promoting critical thinking (2) might require more substantial design changes, as we will discuss in section~\ref{sec:limitations}.
Some of the challenges participants faced in \condbaseline{} could also be addressed with design changes.
For example, several browser extensions that integrate \acp{LLM} with content (for example, by clicking a portion of content to provide it as context to an inline chat window) could make it easier to provide local context.
\section{Implementation}\platform{} is implemented as a plugin for Visual Studio Code, built using TypeScript, React, and Redux\footnote{The \anonymized{(anonymized)} source code for \platform{} is available at \url{https://github.com/anonymizer/EDBook}}.
Content is stored as a \ac{JSON} file with the extension \texttt{.dpage}.

\begin{figure}
    \centering
    \includegraphics[width=\textwidth]{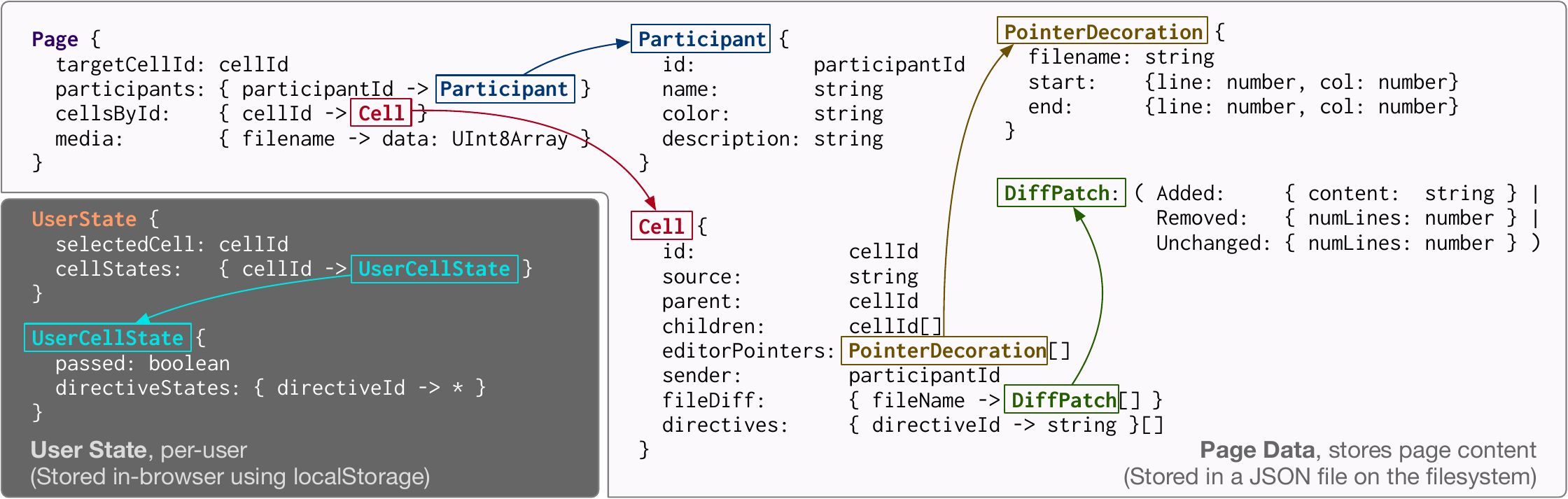}
    \caption{An illustration of how \platform{} content is represented and stored. We differentiate between \emph{page data}---which stores the content of the page and is stored in a file---and \emph{user data}---which represents user-specific information (such as which multiple choice options they selected or which message is currently visible) that is stored in ephemeral local storage.}
    \label{fig:architecture}
\end{figure}

\subsection{Page Format}
Figure~\ref{fig:architecture} illustrates the structure of \platform{} content files (\ac{JSON} data with a \texttt{.dpage} extension).
The format of \platform{} content was inspired by that of Jupyter's \texttt{.ipynb} Notebooks~\cite{kluyver2016jupyter}. 
However, some notable differences with Jupyter's format are (1) a tree structure for cells (rather than lists); (2) a notion of users and provenance to track who sent which message; (3) ``pointers'' that better link narratives with code; (4) code diffs that allow code to change as the narrative advances; (5) education-specific ``directives'' such as multiple-choice questions and coding questions; and (6) separating users' state from the page data.
Every node in the dialog tree contains message text, speaker metadata, code pointers, and directive metadata like multiple choice questions. 

User progress through dialog trees is maintained in browser local storage.
This includes the current node, directive answers, and whether nodes were visited.
Separating mutable user state from immutable page content allows pages to be shared between learners without passing on user-specific data.
It also allows \platforms{} to store user data locally, even for content that is deployed publicly.

\subsection{Syntax, Custom Directives, and Extensibility}
\platform{} documents use Markdown, a popular and simple markup language.
\platform{} uses an extended version of the Markdown syntax that enables custom generic directives, which we use for interactive assessments.
For example, a user can author a multiple-choice question:

\begin{lstlisting}
:::multiple-choice
Which of the following adds up to `2`?
::option[5+5]{feedback="Not quite. That adds to 10"}
::option[1+1]{correct}
::option[2+2]{feedback="Not quite. That adds to 4"}
:::
\end{lstlisting}

Whenever an author updates a message's content, \platform{} scans the message for any custom directives (including multiple-choice and code writing).
It then creates a unique ID for each directive by concatenating the message's ID with the directive type (e.g., \texttt{multiple-choice}) and the index of that directive.
This way, users do not need to specify unique IDs manually~\cite{ericson2020runestone} and these IDs are resilient to wording changes and other small modifications.
This unique directive ID is used as a lookup key to store and recall the user's state for that directive (e.g., which multiple choice options are selected or the user's current code).
Although the \platform{} platform does not have a formal plugin \acs{API}, adding new directives only involves defining a React or JavaScript component that can (1) render \acs{HTML} output, (2) serialize the component state, and (3) subsequently load that serialized state.
Everything else (including tracking multiple directives' states, determining when to render directives, and more) is handled by the \platform{} platform.
In the current implementation, directive states are entirely local and self-contained.
This means, for example, that a multiple-choice question could not render differently depending on a user's response to a different question.

\subsection{Tracking Larger Code Samples, Code Progression, and Incremental Edits}
Every message can have any number of larger associated code samples that get displayed when that cell is selected.
These code samples are stored separately from the cell source.
Authors can write code samples by simply selecting a cell and updating the code (e.g., adding files, removing files, or editing file content).
Because these code samples can be lengthy and are likely to remain unchanged between multiple cells, the \platform{} platform stores code \emph{changes} between cells, rather than the full code.
This helps reduce the file size of the resulting \texttt{.dpage} files.
As Figure~\ref{fig:architecture} illustrates, every message (cell) contains a list of textual diffs.
These represent additions, removals, and unmodified lines relative to prior versions.
Whenever a modification that may change the code is made (e.g., a cell is moved or deleted; or the code associated with a cell is changed), the code state for every cell  is re-constructed and the code diffs are re-derived.

\subsection{Other Implementation Details}
OpenAI's ChatGPT \ac{API} powers contextual \ac{LLM} interactions.
Readers can choose which model to query (GPT-3.5 by default).
When users query the \ac{LLM}, the current dialog context is combined with their question to generate a response.
Constraining \ac{LLM} prompts with dialog context mitigates unhelpful responses.

Uploaded images and other media are converted to binary data (\texttt{UInt8Array}), associated with a filename, and stored as part of the \texttt{.dpage} file (in the \texttt{media} field, as Figure~\ref{fig:architecture} illustrates).
Any deictic code references are stored as additional cell metadata, separate from the cell's source.
\section{Limitations and Future Work}Our design of \platform{} and its evaluation have several limitations that provide opportunities for future work.

\subsection{Encouraging Meaningful Exploration and Reducing Manual Work}

As Figure~\ref{fig:llm_open_ended_design_space} illustrates, \platforms{} exist in a continuum between less constrained tools (e.g., open-ended \acp{LLM}) and more constrained tools (e.g., pre-written linear dialogs).
Although its design enables freeform user queries, the design nudges readers towards the pre-written narrative rather than self-directed exploration.
\platforms{} were designed this way partially because of concerns about the accuracy and usefulness of information generated by \acp{LLM}---we prioritize presenting messages that we can guarantee to be accurate and pedagogically productive.
As \ac{LLM} capabilities (and our trust in them) continue to improve, future systems could instead consider nudging readers towards open-ended interactions and emphasizing \ac{LLM}-generated content.
Future work could also involve fine-tuning \acp{LLM} to improve their ability to generate accurate and contextually relevant outputs that align with specific learning goals.
Further, the \platform{} design could also be adapted to use \acp{LLM} to generate additional assessments and interactive content.

Further, although content authors can use \acp{LLM} to pre-generate content, we found that authoring branched narratives and coding challenges requires significant effort.
Even if authors use \acp{LLM}, it can take significant effort to engineer prompts that produce high-quality content.
Semi-automating content generation through AI could make authoring more scalable.
However, safety mechanisms will also be needed to address potential issues around bias, misinformation, and harmful content from \acp{LLM}.

\subsection{Collaborative Features}
In its current design, readers navigate \platform{} content individually.
However, readers would likely benefit and learn from other readers' inquiries and interactions with \platform{} content.
For example, future versions could give readers the option to share their open-ended interactions with other learners.
These shared user-generated narratives could be curated to ensure that their content is appropriate, accurate, and high-quality.
Reputation systems could surface high-quality user contributions and community voting or experts could identify helpful perspectives.
User-generated content might even be able to ultimately replace the original dialog-tree to create community-written narratives that continue to evolve and improve over time.

\subsection{Generalizing Beyond Programming}
Many aspects of the \platform{} platform's design could apply to domains beyond programming.
For example, dialogs could guide readers through advanced \acp{UI} or creative and open-ended domains.
To start, future work could replace \platform{}'s code panel with applications that are appropriate for the given domain.
With different prompting strategies, users could author customized tutorials for their own domain of work.
\label{sec:limitations}
\section{Conclusion}

Programming education at scale remains a vital challenge.
Traditional materials often lack interactivity, guidance, and engagement.
Although learning with open-ended \acp{LLM} have many benefits, it also has certain negative impacts such as unreliable responses and a lack of a clear learning goal.
This work presents \platform{}, a platform that combines goal-aligned dialog trees with open-ended \ac{LLM} interactions.
We describe the design and implementation of tree-based dialog management, goal alignment techniques, and personalized programming exercises.
Our user study demonstrates several key benefits when using \platforms{} versus traditional materials.
Navigating the balance between open-ended \ac{LLM} interactions and pedagogical goals is still an ongoing research challenge.
Learners benefited from the conversational agents to pose questions, but structured guidance was needed to maintain educational focus.
Our work contributes to the field of human-centric \ac{AI} in programming education by demonstrating promising methodologies.
We see great potential for dialogic learning at scale through \ac{AI}.
The design insights from \platform{} point towards more engaging, responsive, and human-aligned programming education with \ac{AI}.

\bibliographystyle{acm}
\bibliography{references}
\end{document}